\documentstyle[amsfonts,12pt]{article}
\def\one{1\hskip-.37em 1}

\def\half{\textstyle{\frac{1}{2}}}

\def\H{{\cal H}}
\def\D{{\cal D}}
\def\K{{\cal K}}
\def\l{\lambda}
\def\n{\nonumber}
\def\No{\noindent}
\def\R{{\mathbb R}}
\def\p{\phi}
\def\b{\begin{eqnarray*}}                  
\def\e{\end{eqnarray*}}                    
\def\bn{\begin{eqnarray}}                  
\def\en{\end{eqnarray}}                    
\def\tint{\textstyle\int}                 
\def\<{\langle}

\def\ra{\rightarrow}

\def\vt{|{\o\eta}\>}
\def\tv{\<{\o\eta}|}
\def\o{\overline}
\def\a{\alpha}
\def\H{{\cal H}}
\def\D{{\cal D}}
\def\K{{\cal K}}
\def\l{\lambda}
\def\n{\nonumber}
\def\p{\phi}
\def\b{\begin{eqnarray*}}                  
\def\e{\end{eqnarray*}}                    
\def\bn{\begin{eqnarray}}                  
\def\en{\end{eqnarray}}                    
\def\tint{\textstyle\int}                 
\def\<{\langle}
\def\>{\rangle}

\def\E{{\mathbb E}\,}
\def\{{\lbrace}
\def\}{\rbrace}

\title{Product Representations and the \\ Quantization of Constrained Systems}
\author{John R. Klauder\\
Departments of Physics and Mathematics\\
University of Florida\\
Gainesville, Fl  32611\\
Faddeev {\it Festschrift},\\ Steklov Mathematical Institute Proceedings}
\date{}                               
\begin{document}
\maketitle
\begin{abstract}
We study special systems with infinitely many degrees of freedom with regard 
to dynamical evolution and fulfillment of constraint conditions. Attention 
is focused on establishing a meaningful functional framework, and for that 
purpose, coherent states and reproducing kernel techniques are heavily 
exploited. Several examples are given.
\end{abstract}
\section{Introduction}
Generally speaking, the quantum theory of infinitely many degrees of 
freedom (i.e., quantum field theory) exhibits a number of complications. 
However, 
the quantum theory of ``product systems'', also involving infinitely many 
degrees of freedom, is especially simple, and such examples can serve as 
training models for more complicated cases. Initially, one starts with a 
{\it basic system} composed of a finite number of degrees of freedom. To 
be specific, let us say standard canonical degrees of freedom, which is 
the case we study. Subsequently, one adjoins an infinite number of 
{\it identical and independent} basic systems to build a model with an 
infinite number of degrees of freedom. The quantum theory of such systems 
involves (tensor) {\it product representations} of the basic operators, 
and generally needs only an energy scale renormalization. (Some aspects 
of product representations may be found in \cite{kkk}.) On the other 
hand, such models---just like far more complicated examples---require 
that the field-operator representation be carefully chosen with the 
dynamics in mind.
In the present paper we extend the discussion of such models to include 
constraints of a rather general nature and do so in such a way that the 
original product representation is maintained. 
We start with a discussion of basic classical models for finitely many 
degrees of freedom and then illustrate the extension of these classical 
models to infinitely many degrees of freedom in a manner that preserves 
the equality and the independence of each of the basic units that make 
up the infinite system.

\subsection{Classical formulation}
>From a classical point of view, let us start with a $J$ degree of freedom 
model, $1\le J<\infty$, and a classical action given by
  \bn I=\tint[\half(p{\cdot}{\dot q}-q{\cdot}{\dot p})-H(p,q)-
\l^\a(t)\p_\a(p,q)]\,dt\;,  \en
where $p=\{p^j\}_{j=1}^J$ and $q=\{q^j\}_{j=1}^J$ are dynamical 
variables, $p{\cdot}{\dot q}\equiv\Sigma_1^J p^j{\dot q}^j$, etc.,  
and $\{\l^{\a}\}^A_{\a=1}$ denote Lagrange multipliers. We next extend 
this model to $N$ identical and independent copies, $N<\infty$, leading 
to $NJ<\infty$ degrees of freedom. This procedure gives rise to the 
classical action
  \bn I_{(N)}=\Sigma_{n=1}^N\tint[\half(p_n{\cdot}{\dot q}_n-q_n{\cdot}
{\dot p}_n)-H(p_n,q_n)-\l_n^\a(t)\p_\a(p_n,q_n)]\,dt\;,  \en
an expression which exhibits an {\it interchange symmetry} 
$(p_n,q_n)\longleftrightarrow(p_m,q_m)$, $1\le n,m\le N$, for any pairs 
$m$ and $n$. Here $H(p,q)$ denotes the classical Hamiltonian and 
$\{\phi_\a(p,q)\}_{\a=1}^A$ the constraints. 
So long as $N<\infty$ this generalization is straightforward. However, 
things become much more interesting when $N\ra\infty$. Our ultimate 
interest lies in studying the quantum theory of the classical theory 
characterized by the classical action
 \bn I_{(\infty)}=\Sigma_{n=1}^\infty\tint[\half(p_n{\cdot}
{\dot q}_n-q_n{\cdot}{\dot p}_n)-H(p_n,q_n)-\l_n^\a(t)\p_\a(p_n,q_n)]
\,dt\;.  \en
Already at the classical level, in order for this expression to make 
sense, it is necessary that
\bn I_n\equiv\tint[\half(p_n{\cdot}{\dot q}_n-q_n{\cdot}{\dot p}_n)-
H(p_n,q_n)-\l_n^\a(t)\p_\a(p_n,q_n)]\,dt
\en
vanish as $n\ra\infty$. Without loss of generality, we may assume this 
will occur provided $p_n\ra0$ and $q_n\ra0$ combined with the condition 
that $H$ and all $\p_\a$ are continuous functions and that $H(0,0)=0$ as 
well as $\p_\a(0,0)=0$. In that case, as $p_n\ra0$ and $q_n\ra0$, then 
$I_n\ra0$. However, that behavior is not quite enough since it does not 
automatically imply convergence of the series in (3). We do not pursue the 
classical story further but simply
assume that $I_n\ra 0$ sufficiently rapidly so that (3) converges absolutely.
The resultant sequences characterize the domain of the classical theory. 
For instance, some examples may satisfy the criterion $\Sigma_{n=1}^\infty
\,[\,\Sigma_{j=1}^J(|p^j_n|+|q^j_n|)\,]<\infty$. 

Observe that we can also recover $I_{(N)}$ from $I_{(\infty)}$ merely by 
setting $p_n\equiv0$ and
$q_n\equiv0$ for all $n>N$. In this sense we also have the rule that
  \bn I_{(\infty)} = \lim_{N\ra\infty}\,I_{(N)} \en
provided $I_n\ra0$ in a suitable fashion, which, in turn, will hold if 
$p_n\ra0$ and $q_n\ra0$ in an appropriate manner.

Our goal is to discuss the quantum theory of the models classically 
described by (3). Several simple examples are discussed in Section 3. 
\section{Quantum Theory}
\subsection{Basic systems}
Our goal here is to find a meaningful functional formalism for the quantum 
theories involved, including dynamics and constraints.
In our quantum analysis, we shall exploit canonical coherent states and 
for that purpose we choose (with $\hbar=1$)
  \bn |p,q\>\equiv\exp(ip{\cdot Q}-iq{\cdot P})\,|\eta\>  \en
expressed in conventional terms and where the fiducial vector $|\eta\>$ is, 
for the present, a general unit vector.
For any $|\eta\>$, such coherent states admit a resolution of unity in 
the form
\bn \one=\tint|p,q\>\<p,q|\,d\mu(p,q)\;,\hskip1cm d\mu(p,q)=\Pi_{j=1}^J
\,dp^j\,dq^j/2\pi\;, \en
with integration over the entire phase space $\R^{2J}$.
If $\H$ denotes the quantum Hamiltonian operator, then the propagator in 
the coherent-state representation is determined by
  \bn \<p'',q''|\,e^{-i\H T}\,|p',q'\> \;, \en
and this expression may be given a coherent-state path-integral 
representation with no difficulty. Following conventional notation 
\cite{klbs},
 \bn &&\hskip-.4cm\<p'',q''|\,e^{-i\H T}\,|p',q'\>  \n\\
   &&\hskip.6cm=\lim_{\epsilon\ra0}\int\Pi^N_{l=0}\,\<p_{(l+1)},q_{(l+1)}|
(\one-i\epsilon\H)\,|p_{(l)},q_{(l)}\>\,\Pi_{l=1}^Nd\mu(p_{(l)},q_{(l)}) \n\\
&&\hskip.6cm={\cal M}\int e^{i\tint[\half(p{\cdot}{\dot q}-q{\cdot}{\dot p})
- -H(p,q)]\,dt}\,\D p\,\D q\;, \en
the last relation being formal but standard. In making this identification 
we have set $H(p,q)=\<p,q|\H(P,Q)|p,q\>$.

Next, let us temporarily set $\H=0$ and focus on the constraints. To 
introduce quantum constraints, we adopt the {\it projection-operator 
approach} \cite{kla} in which one focuses on the projection operator $\E$ 
onto the physical Hilbert space ${\mathbb H}_{\rm phys}=\E{\mathbb H}$ 
composed of vectors $|\psi\>_{\rm phys}=\E|\psi\>$ for arbitrary 
$|\psi\>\in{\mathbb H}$. It is possible to construct a general $\E$ by a 
linear operation on the set of unitary operations generated by the 
constraints. In particular, if the several self-adjoint operators 
$\Phi_\a(P,Q)$ denote the quantum constraint operators with the property 
that $\Sigma\Phi_\a(P,Q)^2$ is essentially self adjoint, then there 
exists \cite{kl2} a linear operation that is {\it independent of the 
specific constraint operators themselves}, denoted by an integral with 
measure $R(\l)$, and such that
  \bn \int{\mathbb T}\exp[-i\tint_t^{t+\epsilon}\,\l^\a(s)\,\Phi_\a(P,Q)
\,ds]\,\D R(\l)=\E(\!\!(\Sigma\Phi_\a(P,Q)^2\le\delta(\hbar)^2)\!\!)\;.  \en
Here ${\mathbb T}$ denotes time ordering, $\epsilon>0$, and 
$\delta(\hbar)^2>0$ denotes a suitable, and possibly provisional, 
precision with which the constraints are enforced. 

A few examples will illustrate how this concept may be used. If 
$\{\Phi_\a\}$ denotes operators with discrete spectra, say angular 
momentum operators $J_k$, $k\in\{1,2,3\}$, then $\delta(\hbar)^2\le 
\hbar^2/10$ ensures that $\E=\E (\Sigma J_k^2=0)$. If $\{\Phi_\a\}$ 
denotes second-class constraints, say $\Phi_1=P$ and $\Phi_2=Q$, then 
$\delta(\hbar)^2=\hbar$ ensures that $\E=\E (P^2+Q^2\le\hbar)=|0\>\<0|$, 
the projection operator onto the oscillator ground state. If $\{\Phi_\a\}$ 
denotes an operator with zero in its continuous spectrum, say $\Phi_1=P$, 
then $\E=\E (P^2\le\delta^2)$ and $\delta^2>0$ can be chosen arbitrarily 
small, e.g., $\delta^2=10^{-100}$. For all practical purposes it is not 
necessary that $\delta\ra0$; however, that limit can also be incorporated 
with a possible change of the Hilbert space involved. 

The mechanism for a possible change of Hilbert space arises by a {\it 
reduction of the reproducing kernel} \cite{kla}. In particular, if 
$\K(p'',q'';p',q')\equiv\<p'',q''|p',q'\>$ denotes the reproducing 
kernel \cite{kl4} for the full Hilbert space of the unconstrained system, 
then $\K_{\E}(p'',q'';p',q')\equiv\<p'',q''|\E|p',q'\>$ denotes the 
reproducing kernel for the (provisional) physical Hilbert space 
appropriate to the constrained system. To illustrate a reduction of such 
expressions, we set $J=1$ and focus on the example
 \bn &&\hskip-.6cm\K_{\E}(p'',q'';p',q')=\<p'',q''|\E(P^2\le\delta^2)
|p',q'\>\;, \n\\
     &&\hskip1cm =e^{-i\half(p''q''-p'q')}\int_{-\delta}^\delta
\eta(k-p'')^*\,e^{ik(q''-q')}\,\eta(k-p')\,dk\;. \en
As $\delta\ra0$ this expression vanishes, but if we first divide by 
$\delta$ before taking the limit, we can generate a positive-definite 
function which, if continuous, characterizes a new Hilbert space, the true 
${\mathbb H}_{\rm phys}$. In particular, let us assume that $\eta(k)$ is 
a continuous function, multiply by $1/2\delta$, and take the limit 
$\delta\ra0$, leading to the result $e^{-ip''q''/2}\eta(-p'')^*
\eta(-p')e^{ip'q'/2}$. The resultant expression is a reproducing kernel 
for a one-dimensional Hilbert space. 

In more abstract terms, and in cases where the dependence of $\K_\E$ on 
$\delta$ is less clear, we can proceed as follows. Let
  \bn  W\equiv \limsup_{(p,q)\in\R^{2J}}\; \<p,q|\E|p,q\>\;,  \en
for which, provided $\E\not\equiv0$, $W>0$. To show that $W$ is positive, 
we observe that
\bn 0\le|\<p,q|\E|r,s\>|^2\le\< p,q|\E|p,q\>\<r,s|\E|r,s\>\le W^2\;. \en
If $W=0$, then it would follow that $\<p,q|\E|r,s\>=0$ for all arguments, 
which can only hold if $\E=0$, contrary to our assumption. Armed with $W$ 
we next define
  \bn \K_W(p'',q'';p',q')\equiv W^{-1}\<p'',q''|\E|p',q'\>\;. \en
Note that $|\K_W|\le1$. We first observe that $\K_W$ corresponds to a new 
(simply rescaled) reproducing kernel for which every element of the 
associated Hilbert space is already a member of the space determined 
by $\K_\E$. To reduce this expression we simply take the limit 
$\delta\ra0$, namely,
 \bn \K_R(p'',q'';p',q')\equiv\lim_{\delta\ra0}\K_W(p'',q'';p',q')\;. \en
If the result of this $\delta$-limiting procedure exists and is continuous, 
then the result is a reproducing kernel for the ultimate physical Hilbert 
space. As we have already seen, the dimensionality of the Hilbert space 
can change dramatically in this limit and, moreover, some of the 
variables may no longer be relevant. Such a procedure may also change 
the measure (if any) by which the inner product in the new space may be 
evaluated. 
 
We can combine constraints with a nonvanishing Hamiltonian by the 
observation that 
\bn  &&\hskip-1.3cm\<p'',q''|\E \,e^{-i(\E\!\H\E) T}\,\E|p',q'\>  \n \\
    &&\hskip-1cm=\lim_{\epsilon\ra0}\;\<p'',q''|\E e^{-i\H\epsilon}
\E e^{-i\H\epsilon}\E\cdots\E e^{-i\H\epsilon}\E|p',q'\> \n\\
&&\hskip-1cm={\cal M}\int e^{i\tint[\half(p\cdot {\dot q}-q\cdot{\dot p})
- -H(p,q)-\l^\a\phi_\a(p,q)]\,dt}\,\D p\,\D q\,\D E(\l)\;, \en
where $\phi_\a(p,q)=\<p,q|\Phi_\a(P,Q)|p,q\>$, and $E(\l)$ is a measure, 
based on $R(\l)$, that is designed to introduce the projection operator 
$\E$ at every time slice. When $\E\H=\H\,\E$, then a significant 
simplification occurs. In that case we may make use of the relation
 \bn \E \,e^{-i(\E\!\H\E)T}\,\E = e^{-i\H T}\,\E \en
which holds as an identity. Thus, in this case, 
 it is only necessary to put {\it one} projection operator $\E$ inside 
the matrix elements to achieve the same result. Although it is possible 
to use $E(\l)$ in this latter case as well, it may be easier to use a 
measure $C(\l)$ designed to insert (at least) {\it one} projection 
operator $\E$. In (16), observe how the evolution operator in 
${\mathbb H}_{\rm phys}$, namely $\exp[-i(\E\H\E)T]$, is evaluated in 
terms of matrix elements of vectors in the physical Hilbert space, 
namely $\E|p,q\>$. Such an expression is fully consistent with the 
constraints. For instance, in the case of closed first-class constraints, 
the propagator within the physical Hilbert space (16) is {\it manifestly 
gauge invariant}, provided one has also used a $\delta$-limiting procedure 
if necessary. 

As (16) shows, the propagator within the physical Hilbert space is 
obtained by means of a formal path integral (with a meaningful lattice 
formulation and lattice limit) involving just the original dynamical 
variables and the Lagrange multipliers. No other variables are needed. 
How this quantization procedure for constrained systems relates to other, 
better known procedures is briefly discussed elsewhere \cite{kla}.
\subsection{Finitely many basic systems}
In this subsection we take up the quantization of a classical system 
described by (2) based on our discussion of the quantization of (1). Due 
to the independence of the separate basic systems, this extension is 
straightforward. Let us extend our notation so that now 
$p=\{p_n\}_{n=1}^N$ and $p_n=\{p_n^j\}_{j=1}^J$, etc., and therefore as a 
consequence
  \bn \K_{(N)}(p'',q'';p',q')\equiv\Pi_{n=1}^N\K(p''_n,q''_n;p'_n,q'_n) \en
denotes the reproducing kernel for the $NJ$ degree of freedom system. 
Since the separate reproducing kernels in (18) do not depend on $n$, it 
is evident that $\K_{(N)}$ is invariant under the interchange of variables 
for any pair of independent basic systems just as is the case for the 
classical theory.

Dynamics (without constraints) takes the form
\bn \<p'',q''|\,e^{-i\H_{(N)}T}\,|p',q'\>=\Pi_{n=1}^N\<p''_n,q''_n|
\,e^{-i\H T}\,|p'_n,q'_n\>\;, \en
while the imposition of constraints (without dynamics) leads to
\bn  \<p'',q''|\,\E_{(N)}|p',q'\>=\Pi_{n=1}^N\<p''_n,q''_n|
\,\E|p'_n,q'_n\>\;. \en
Reduction of such reproducing kernels follows the pattern described in the 
previous subsection.
Finally, combining dynamics and constraints generally leads to 
\bn &&\hskip-.5cm\<p'',q''|\E_{(N)}e^{-i(\E_{(N)}\H_{(N)}\E_{(N)}) T}
\,\E_{(N)}|p',q'\> \n\\
&&\hskip2cm=\Pi_{n=1}^N \<p''_n,q''_n|\E \,e^{-i(\E\!\H\E) T}
\,\E|p'_n,q'_n\>\;, \en
or, in the special case that $\E\H=\H\,\E$, to
\bn \<p'',q''|\,e^{-i\H_{(N)}T}\,\E_{(N)}|p',q'\>=\Pi_{n=1}^N\<p''_n,q''_n|
\,e^{-i\H T}\,\E|p'_n,q'_n\>\;. \en
All these expressions exhibit the interchange symmetry inherit in the 
classical system.

\subsection{Infinitely many basic systems}
Due to the elementary structure of product representations, the analysis 
of infinitely many independent basic systems largely involves only a 
study of the limit $N\ra\infty$ in several formulas of the preceding 
section. Naturally, convergence of such limits will be a critical issue.

First, just for kinematics, with neither dynamics nor constraints, we 
require that 
\bn\<p'',q''|p',q'\>=\Pi_{n=1}^\infty\,\<p''_n,q''_n|p'_n,q'_n\> \;,\en
and convergence of the right-hand side dictates what elements 
$p=\{p_n\}_{n=1}^\infty$ and $q=\{q_n\}_{n=1}^\infty$ may enter on the 
left-hand side. 
The value of zero for this product may be arrived at in two different 
ways: (i) either one (or more) of the factors vanishes, or (ii) every 
factor is nonzero, but the infinite product leads to zero. This latter 
situation is called ``divergence to zero'', and in discussions regarding 
this subject \cite{vne} it is not considered convergence.
To have convergence, and to exclude divergence to zero, we need that
 \bn \Sigma_{n=1}^\infty\,|1-\<p''_n,q''_n|p'_n,q'_n\>|<\infty \;. \en
Since each coherent state is a unit vector, convergence only occurs 
provided that $q''_n-q'_n\ra0$ and $p''_n-p'_n\ra0$. To
preserve interchange symmetry, we require, in turn, that 
$q''_n\ra{\o q}$, $q'_n\ra{\o q}$ and $p''_n\ra{\o p}$, $p'_n\ra{\o p}$, 
where $({\o p},{\o q})\in \R^{2J}$ is arbitrary. Observe that the 
variables $({\o p},{\o q})$ which label the asymptotic dependence 
actually label {\it orthogonal Hilbert spaces}. This statement holds 
because if $({\o p},{\o q})\not=({\o r},{\o s})$, then $|\<{\o p},{\o q}|
{\o r},{\o s}\>|<1$ and the infinite power of this factor yields zero; 
this is an example of divergence to zero. No change of a finite number 
of labels in either the bra or the ket, nor finite linear superpositions 
and arbitrary Cauchy sequences thereafter,  can ever change the vanishing 
result. We deal here with an uncountable number of disjoint reproducing 
kernel Hilbert spaces (save for the zero element). (Stated alternatively, 
if one were to realize the underlying field operators in a common Hilbert 
space, then $({\o p},{\o q})$ would label {\it unitarily inequivalent 
irreducible representations} \cite{kkk}.) At this stage, there is no 
distinguished property that would help us choose which $({\o p},{\o q})$ 
set or which fiducial vector $|\eta\>$ is correct. In fact, that is as 
it should be since we have not specified any particular dynamics. In 
summary, labels for the coherent states are given by the set $T({\o p},
{\o q})\equiv\{(p_n,q_n):p_n\ra{\o p},q_n\ra{\o q}\}$ where convergence 
means that 
\bn \lim_{N\ra\infty}\Pi_{n=N}^\infty\;\<p_n,q_n|{\o p},{\o q}\>=1  \en
or equivalently 
\bn \lim_{N\ra\infty}\Sigma_{n=N}^\infty\; |1-\<p_n,q_n|{\o p},{\o q}\>|
=0\;.  \en  
This criterion applies for any choice of $|\eta\>$, and leads to 
acceptable (possibly $|\eta\>$-dependent) momentum and coordinate 
variable sets $(p,q)=\{p_n,q_n\}_{n=1}^\infty$.

If one adds a modest domain requirement on the fiducial vector, such as 
$\<\eta|(P^2+Q^2)|\eta\><\infty$, then the convergence criterion in (26) 
is equivalent to 
\bn  \Sigma_{n=1}^\infty\,[\;\Sigma_{j=1}^J(|p^j_n-{\o p^j}|+|q^j_n-
{\o q}^j|)\,]<\infty\;,  \en
a relation that captures the allowed sequences $\{p_n,q_n\}_{n=1}^\infty$ 
for a wide class of examples.

Next, let us consider the case of dynamics without constraints. Thus we 
initially study
 \bn \<p'',q''|\,e^{-i\H_{(\infty)} T}\,|p',q'\>=\Pi_{n=1}^\infty
\,\<p''_n,q''_n|\,e^{-i\H T}\,|p'_n,q'_n\>\;. \en In the case of dynamics, 
proper convergence of (28) requires, for acceptable sets $(p,q)\in 
T({\o p},{\o q})$, for some $({\o p},{\o q})$,  that
\bn \Sigma_{n=1}^\infty\,|1-\<p''_n,q''_n|e^{-i\H T}\,|p'_n,q'_n\>|
<\infty \;.  \en
Without loss of generality, we shall assume that ${(\o p},{\o q})=(0,0)$. 
In that case, the criterion (29) requires that $\H|\eta\>=0$. If 
$\H$ has a (partially) discrete spectrum, then $|\eta\>$ may be taken as 
an eigenvector whose eigenvalue has been adjusted to vanish. If $\H$ has 
only a continuous spectrum, then it is not possible to satisfy (29) as it 
stands unless we allow for $n$-dependent fiducial vectors, a modification 
that would destroy interchange symmetry. Since we wish to preserve 
interchange symmetry, we must confine attention to $|\eta\>$ being a 
fixed, normalized eigenvector of $\H$. Observe that this condition links 
kinematics and dynamics, a condition generally regarded as a hallmark of 
infinitely many degrees of freedom (c.f., Haag's Theorem \cite{hag}). 
In other words, in order for $|\eta\>=\Pi_{n=1}^\infty(|\eta\>)_n$ to 
be a unit vector in the full Hilbert space in which the Hamiltonian 
$\H_{(\infty)}$ is a well-defined (and self-adjoint) operator requires 
that $\H|\eta\>=0$. Hence, far from being chosen arbitrarily, $|\eta\>$ 
is now determined to be an eigenvector of $\H$; when $\H\ge0$, we may 
even choose $|\eta\>$ to be one of the ground states. If, as is often 
the case, the ground state is unique, then $|\eta\>$ is fixed. Thus we 
see that the introduction of dynamics has effectively selected the 
fiducial vector $|\eta\>$, as well as the parameters 
$({\o p},{\o q})=(0,0)$, in order that $\H_{(\infty)}$ is a self-adjoint 
operator. For convenience in what follows, we generally restrict attention 
to Hamiltonian operators $\H$ with a purely discrete spectrum.

Next, we consider constraints but no dynamics, a situation which---for 
the moment---restores general values of $({\o p},{\o q})$ and general 
$|\eta\>$ to consideration. As a first approach to this problem, consider
\bn  \<p'',q''|\E_{(\infty)}|p',q'\>=\Pi_{n=1}^\infty\<p''_n,q''_n|\E
|p'_n,q'_n\>\;,  \en
where each argument set is a member of $T({\o p},{\o q})$. 
By Schwarz's inequality the right-hand side of (30) is a product of 
factors each of which is at most unity in magnitude.  Therefore, as 
it stands, in order for  this product to converge (and
not diverge to zero), it is necessary, for some $({\o p},{\o q})$ and 
$|\eta\>$, that $\E|{\o p},{\o q}\>=|{\o p},{\o q}\>$, namely that
the vector $|{\o p},{\o q}\>$ {\it already} belongs entirely to the 
physical Hilbert space. If this condition is fulfilled, then (30) defines 
a valid reproducing kernel on the physical Hilbert space for infinitely 
many degrees of freedom. On the other hand, this condition is a very 
strong restriction. We shall next see how we can significantly relax 
this requirement. 

Suppose, as dictated by the future dynamics, that $({\o p},{\o q})=(0,0)$ 
and that $|\eta\>$ satisfies  $\E|\eta\>\ne|\eta\>$. Several situations 
are then possible. If $\E|\eta\>=0$, then the vector $|\eta\>$ lies 
entirely in the {\it un}physical Hilbert space and $S=\<\eta|\E|\eta\> =0$. 
This property simply means that the chosen eigenvector of $\H$ is 
{\it incompatible with} $\E$. We cannot change $\H$ or $\E$, but we can 
change the fiducial vector. Hence, we introduce a new and distinct 
fiducial vector $\vt$ such that $\E\vt\not=0$ and thus 
${\o S}=\tv\E\vt>0$. It may even be appropriate to choose $|{\o\eta}\>=
|{\o p},{\o q}\>$ for some $({\o p},{\o q})\ne(0,0)$. Further 
conditions on $\vt$ will appear below. 

Armed with ${\o S}$ we introduce the rescaled reproducing kernel
 \bn {\o\K}_R(p'',q'';p',q')=\Pi_{n=1}^\infty\,[\;{\o S}^{\,-1}
\<p''_n,q''_n|\E|p'_n,q'_n\>] \;, \en
where in this expression $|p,q\>$ is defined as in (6) with $\vt$ 
used in place of $|\eta\>$, and also such that $(p_n,q_n)\ra(0,0)$. 
In this language, ${\o S}=1$ corresponds to the case of (30) where 
$\E\vt=\vt$. 

Finally, we turn to the case of dynamics plus constraints which will 
lead to additional conditions on $\vt$. The putative propagator reads
\bn  K(p'',q'',T;p',q',0)=\Pi_{n=1}^\infty\,[\;{\o S}^{\,-1}\<p''_n,q''_n|
\E\, e^{-i(\E\!{\o\H}\E)T}\,\E|p'_n,q'_n\>]\;, \en
where ${\o\H}=\H-{\o E}$, with $\o E$ to be fixed. In order for this 
product to converge it is necessary that
 \bn \E \,e^{-i(\E\!{\o\H}\E)T}\,\E\vt=\E\vt\;.  \en
Assume that ${\o\H}_\E\equiv \E{\o\H}\E$ is self adjoint, with a 
discrete spectrum, and let $|\xi_l\>$, $l=0,1,2,\ldots$, be a complete 
orthonormal set of eigenvectors that satisfy ${\o\H}_\E|\xi_l\>=
\E{\o\H}\E|\xi_l\> =\sigma_l|\xi_l\>$, with $\sigma_0\le\sigma_1
\le\sigma_2\cdots$. Choose $\vt$ such that $\E\vt=c|\xi_p\>$, $c\ne0$, 
for the least $p$ value, and then choose ${\o E}$ so that $\sigma_p=0$. 
If ${\o E}\ne0$, then this last condition has involved an infinite 
renormalization of the energy. With all conditions satisfied if follows 
that (33) holds and (32) determines the dynamics and constraints together.

The situation is simpler if $\H\,\E=\E\H$, and in that case it is 
sufficient to consider
\bn K(p'',q'',T;p',q',0)=\Pi_{n=1}^\infty\,[\;{\o S}^{\,-1}\<p''_n,q''_n|
\,e^{-i{\o\H} T}\,\E|p'_n,q'_n\>]\;, \en
where each $(p_n,q_n)\ra(0,0)$, ${\o S}=\tv\E\vt>0$, and ${\o\H}=\H-{\o E}$. 
Convergence of this expression requires, along with $\E\vt\ne0$, that
\bn \E\vt=e^{-i{\o\H}T}\,\E\vt=\E e^{-i{\o\H}T}\,\vt\;,  \en
which implies that ${\o\H}\vt=0$. Let ${\o\H}|\zeta_l\>=\mu_l|\zeta_l\>$, 
$l=0,1,2,\ldots$, $\mu_0\le\mu_1\le\mu_2\cdots$, and set $\vt=|\zeta_r\>$ 
for the least $r$ such that $\<\zeta_r|\E|\zeta_r\>>0$, adjusting ${\o E}$ 
to ensure that $\mu_r=0$. In this case the product in (34) converges to 
an acceptable propagator.

\section{Examples}
We illustrate some of the concepts in the previous sections with several 
examples. In order to do so we shall give the classical action for the 
basic system and then present the propagator on the physical Hilbert space 
in the form of a suitable coherent-state functional. In our examples we 
shall exclusively use the harmonic oscillator ground state (for unit 
angular frequency) as the fiducial vector. Thus we deal with the canonical 
coherent states in the so-called holomorphic representation; for notation 
consult, e.g.,  \cite{klbs}. For clarity, we only present simple and 
explicitly soluble examples.  \vskip.3cm
  
\No{\bf Example 1.} Choose $J=A=1$ and the classical action
\bn  I=\tint[\half(p{\dot q}-q{\dot p})-\l(p^2+q^2)]\,dt\;,  \en
which has a vanishing Hamiltonian. In this case 
\bn \E=\E(\!\!((P^2+Q^2)^2\le\hbar^2)\!\!)=\E(\!\!((P^2+Q^2)
\le\hbar)\!\!)=|0\>\<0|\;, \en
namely the projection operator onto the harmonic oscillator ground state.
In terms of the complex variable $z\equiv (q+ip)/\sqrt{2}$ and coherent states
$|z\>\equiv|p,q\>$ and $\<z|\equiv\<p,q|$, we determine that
\bn  \<z''|z'\> = \exp(-\half|z''|^2+z''^*z'-\half|z'|^2) \;,\en
and that 
\bn  \<z''|\E|z'\>=\exp[-\half(|z''|^2+|z'|^2)]\;. \en
Observe that $\<0|\E|0\>=1$ so that the propagator for an infinite 
product system satisfying the constraints is given by
\bn  K(z'',T;z',0)=\Pi_{n=1}^\infty\;e^{-\half(|z''_n|^2+|z'_n|^2)}\;. \en
Convergence of this expression requires for each argument that 
$\Sigma_{n=1}^\infty\,|z_n|^2<\infty$. Observe that the physical 
Hilbert space is one dimensional for all $N$, $1\le N\le\infty$.\vskip.3cm

\No{\bf Example 2.} Let $J=1$, $A=2$ and choose 
\bn  I=\tint[\half(p{\dot q}-q{\dot p})-\half(p^2+q^2)-\l^1 p-\l^2 q]
\,dt\;.  \en
Here 
\bn  \E=\E(\!\!(P^2+Q^2\le \hbar)\!\!)=|0\>\<0| \en
again. With $\H=\half(P^2+Q^2)$, and after adjusting for the zero-point 
energy, the solution is identical to Example 1 and given by (40).\vskip.3cm

\No{\bf Example 3.} Again $J=A=1$, and consider
\bn  \tint[\half(p{\dot q}-q{\dot p})-\half(p^2+q^2)-\l(p^2+q^2-2)]\,dt 
\;. \en
This expression fails to satisfy the conditions following (4), so we must 
already make an energy
renormalization and instead choose
\bn  I=\tint[\half(p{\dot q}-q{\dot p})-\half(p^2+q^2-2)-\l(p^2+q^2-2)]\,dt 
\;, \en
where, e.g., $(p_n,q_n)\ra(1,1)$ classically, i.e,  $z_n\ra\sqrt{i}$.
In this case, 
\bn &&\hskip-.45cm\E=\E(\!\!(P^2+Q^2=3\hbar)\!\!)\n\\
&&=\E(\!\!(P^2+Q^2\le3\hbar)\!\!)-\E(\!\!(P^2+Q^2\le\hbar)\!\!)\n\\
&&=|1\>\<1|\;, \en 
where $|1\>$ denotes the first excited state of the harmonic oscillator. 
It follows, therefore, that
\bn &&\<z''|z'\>=\exp(-\half|z''|^2+z''^*z'-\half|z'|^2)\;,\n\\
 &&\<z''|\E|z'\>=\exp[-\half(|z''|^2+|z'|^2)]\;z''^*z'\;. \en
In this case the chosen fiducial vector---the harmonic oscillator 
ground state---is incompatible with the constraint condition, namely 
$\E|0\>=0$. Thus we need to change the fiducial vector, and for that 
purpose we choose $|{\o\eta}\>=|z=\sqrt{i}\>\equiv|\sqrt{i}\>$.  Next, 
we set ${\o \H}=\H-{\o E}=\H-1$, then ${\o \H}|1\>=0$ and $\E|1\>=|1\>$. 
Since $S=\<\sqrt{i}|\E|\sqrt{i}\>=1/e$, the
propagator coupled with the constraints, following (34), is given by
\bn K(z'',T;z',0)=\Pi_{n=1}^\infty\,\exp[-\half(|z''_n|^2+|z'_n|^2-2)]
\;z''^*_nz'_n\;, \en
an expression which describes a valid propagator on the one-dimensional 
physical Hilbert space. In the present case convergence means that 
$\Sigma_{n=1}^\infty|z_n-\sqrt{i}|<\infty$.  \vskip.3cm

\No{\bf Example 4.} Here $J=A=3$ and 
\bn I=\tint[\half(p{\cdot}{\dot q}-q{\cdot}{\dot p})-
\half(p^2+q^2)-\l^j(p\wedge q)_j]\,dt\;. \en
In this case the three constraints are the angular momentum generators and 
$\E=\E(\Sigma_{j=1}^3 J_j^2=0)$, i.e., a projection operator onto the
spherically 
symmetric subspace. For convenience, let us introduce the notation 
${\bf z}=(z^1,z^2,z^3)$ for the three components, and for two such vectors 
let ${\bf z}{\cdot}{\bf w}\equiv\sum_{j=1}^3 z^jw^j$ and $({\bf z})^2
\equiv\sum_{j=1}^3 (z^j)^2$, etc. Then the reproducing kernel for the full 
Hilbert space is given by
\bn \<z''|z'\>=\exp[\Sigma_{n=1}^\infty(-\half|{\bf z}''_n|^2+
{\bf z}''^*_n{\cdot}{\bf z'}_n-\half|{\bf z}'_n|^2)]\;.  \en
After a modest computation the reproducing kernel for the physical 
Hilbert space is given by
\bn \<z''|\E|z'\>=e^{-\half\Sigma_{n=1}^\infty(|{\bf z}''_n|^2+
|{\bf z}'_n|^2)}\,\prod_{n=1}^\infty\,\sum_{m=0}^\infty
\frac{[({\bf z}''^*_n)^2({\bf z}'_n)^2]^m}{(2m+1)!}\;. \en
Inclusion of the Hamiltonian follows simply be the change 
${\bf z'}_n\ra e^{-iT}\,{\bf z}'_n$, and leads to 
\bn \<z''|\,e^{-i\H T}\,\E|z'\>=e^{-\half\Sigma_{n=1}^\infty
(|{\bf z}''_n|^2+|{\bf z}'_n|^2)}\,\prod_{n=1}^\infty\,
\sum_{m=0}^\infty\frac{[({\bf z}''^*_n)^2({\bf z}'_n)^2]^m}{(2m+1)!}
\,e^{-i2mT}\,.  \en
We observe that with the constraints in force, the energy spectrum is 
$E_m=2m$ rather than $E_m=m$ which applies to the unconstrained 
oscillators. Convergence of this expression requires that each sequence 
$\{{\bf z}_n\}$ satisfy 
\bn  \Sigma_{n=1}^\infty|{\bf z}_n|^2<\infty\;.  \en

\section*{Dedication}
It is a pleasure to dedicate this article to the 65th birthday of 
Ludwig Faddeev. His contributions, in a wide range of scientific fields, 
already place him in the Pantheon of Truly Great Scientists. May he 
continue to enlighten us all for many years to come.

\section*{Acknowledgements}
Jan Govaerts and Sergei Shabanov are both thanked for their 
continued interest in applying projection operator techniques to study 
quantum constrained systems.


\begin{thebibliography}{99}
\bibitem{kkk} J.R. Klauder and J. McKenna, J. Math. Phys. {\bf 6}, 68 (1965);
J. R. Klauder, J. McKenna, and E.J. Woods, J. Math. Phys. {\bf 7}, 822 (1966).
\bibitem{klbs} J.R. Klauder and B.-S. Skagerstam, ``Coherent States'', 
(World Scientific, Singapore, 1985).
\bibitem{kla} J.R. Klauder, Ann. Phys. {\bf 254}, 419 (1997). See also: S. 
Shabanov, in ``Path Integrals: Dubna `96'', Eds. V.S. Yarunin and M.A. 
Smondyrev, (Publishing Department, Joint Institute for Nuclear Research, 
Dubna, Russia), p. 133. 
\bibitem{kl2} J.R. Klauder, in preparation.
\bibitem{kl4} N. Aronszajn, Proc. Cambridge Phil. Soc. {\bf 35}, 133 (1943); 
Trans. Am. Math. Soc. {\bf 68}, 337 (1950).
\bibitem{vne} J. von Neumann, Compositio Math. {\bf 6}, 1 (1938).
\bibitem{hag} R. Haag, Kgl. Danske Videnskab. Seskab, Mat.-fys. Medd. 
{\bf 29}, 12 (1955).



\end{thebibliography}
\end{document}